\documentclass{emulateapj}
\usepackage{aas_macros}
\usepackage{epstopdf}
\usepackage{epsfig}
\usepackage{natbib}
\usepackage{float}
\usepackage{gensymb}
\usepackage{amsmath}
\usepackage{lipsum}% http://ctan.org/pkg/lipsum
\usepackage{times}
\usepackage{comment}
\usepackage{color}
\usepackage{multirow}
\usepackage{soul}
\usepackage{CJKutf8}            % Chinese name

\definecolor{purple}{RGB}{160,32,240}

\definecolor{purple2}{RGB}{120,72,240}

\newcommand{\peteradd}[1]{\textcolor{purple2}{#1}}

\begin{document}

\title{Emission line ratios for the Circumgalactic Medium and the ``Bimodal" Nature of Galaxies}

\author{Huanian Zhang \begin{CJK*}{UTF8}{gkai}(张华年)\end{CJK*}\altaffilmark{1}, Dennis Zaritsky\altaffilmark{1}, Jessica Werk\altaffilmark{2}, and Peter Behroozi\altaffilmark{1}}
\altaffiltext{1}{Steward Observatory, University of Arizona, Tucson, AZ 85719, USA; fantasyzhn@email.arizona.edu}
\altaffiltext{2}{Department of Astronomy, University of Washington, Seattle, WA 98195,  USA}

\begin{abstract}
We find significantly different diagnostic emission line ratios for the circumgalactic gas associated with 
galaxies of stellar masses above and below $10^{10.4}$ M$_\odot$ using SDSS spectroscopy. 
Specifically, in a sample of 17,393 galaxies, intersected by 18,535 lines of sight at projected radii
between 10 and 50 kpc, we  stack measured fluxes  
for nebular strong emission lines, 
[O {\small III}] $\lambda$5007, H$\alpha$ and [N {\small II}] $\lambda6583$, and find that the gas surrounding the lower mass galaxies exhibits similar line ratios to those of gas ionized by star formation and that surrounding the higher mass galaxies similar to those of gas ionized by AGN or shocks. 
This finding highlights yet another characteristic of galaxies that is distinctly different above and below this stellar mass threshold, but one that is more closely connected to the gas accretion behavior hypothesized to be responsible for this dichotomy. 

\end{abstract}


\section{Introduction}

The manner in which gas accretes onto galaxies, and therefore the initial state of the gas observed in the circumgalactic medium, is hypothesized to be distinctly different in high and low mass galaxies \citep{birnboim,keres}. The physical distinction centers on whether the gas is shocked as it is accreted onto the dark matter galaxy halo, as originally envisioned by \cite{white}, or directly accreted onto the central galaxy. The two modes are broadly referred to as ``hot'' and ``cold'' modes of accretion, respectively. 
The different behavior is expected to lead to distinguishing observable galaxy characteristics \citep{keres,dekel}. Many of the key, broad differences among galaxies, such as color, morphology, and current star formation rate, have long been known to track the mass of the galaxy \citep{roberts} and that connection has been unambiguously demonstrated using data from the Sloan Digital Sky Survey \citep{blanton,kauffmann_mass}. In particular, \cite{kauffmann_mass} found that  galaxies separate into ``two distinct families" and that the break in galaxy properties happens relatively sharply at a stellar mass, $M_*$, of $\sim 3\times10^{10}$ M$_\odot$, which corresponds closely to the transition scale for accretion behavior identified in numerical simulations, $M_* = 10^{10.4}$ M$_\odot$ \citep{keres}. 

Since these studies, there has been an explosion in studies of the circumgalactic medium \citep[see][for a review with an extensive list of references]{tumlinson}. The emerging picture is far more complicated, including gas recycling, the contribution of gas from galaxy mergers, and even the accretion of gas processed previously in other galaxies \citep{ford,angles-alcazar}. There are even proposals that the current state of the gas may reflect a previous state of the central galaxy \citep[for example, previous AGN activity;][]{oppenheimer}. While no simple picture will capture all of this richness, there continue to be predictions that the properties of the circumgalactic medium should depend on galaxy mass.

To bridge the gap between the measured central properties of galaxies and the theoretical modeling of gas accretion, 
we investigate whether the observed circumgalactic gas also differs significantly for low and high mass galaxies. 
Diagnostic emission line ratios, like those used in the BPT diagram \citep{bpt}, provide guidance on the ionizing source. Those authors identified regions in line ratio diagrams indicative of excitation by normal star formation (H {\small II} regions), active galactic nuclei (AGN, which are power-law spectrum sources), shock heating, and hot stellar remnants (planetary nebulae). Subsequent studies that focused on the use of such line ratios to interpret galaxy spectra focused on distinguishing between the two expected dominant sources of excitation, star formation and AGN \citep[{e.g.}][]{vo,kewley,kauffmann_agn}.

Our observational challenge is to detect the line emission from the circumgalactic medium (CGM) necessary to construct the diagnostic ratios.
Stacking  thousands to millions of spectra obtained from galaxy redshift surveys has proven to provide an avenue for studying difficult to reach spectral features in a variety of contexts \citep[e.g.][]{steidel2010,menard2011,bordoloi2011,zhu2013a,zhu2013b,Werk2014,werk16,croft2016,croft2018, prochaska2017,lan2018,joshi2018}.
We recently applied the stacking technique to SDSS spectra to uncover emission lines originating in the CGM of low redshift galaxies \citep[][hereafter, Papers I and II]{zhang2016,zhang2018}. These two studies present results on the radial distribution of H$\alpha + $[N {\small II}] emission from the halos of normal galaxies and their neighbors to projected radii  beyond 100 kpc. 
We continue our exploration of this gas using spectral stacking by now measuring additional recombination lines in an effort to constrain the physical state of the gas  in the warm ionized halos of low and high mass galaxies using standard diagnostic line ratios.
To evaluate distances, we adopt standard cosmological parameters $\Omega_m$ = 0.3, $\Omega_\Lambda =$ 0.7, $\Omega_k$ = 0 and the dimensionless Hubble constant $h = $ 0.7.\\

\section{Data Analysis}
We follow the 
approach developed in Papers I and II, but focus on the  
 CGM of host galaxies by considering only sightlines with projected separations $\le$ 50 kpc (see Paper II for a discussion of how neighboring halos begin to dominate the integrated emission profile at larger projected radii).
We obtain galaxy spectra from the Sloan Digital Sky Survey Data Releases \cite[SDSS DR12]{SDSS12} and classify galaxies that meet our  criteria in redshift (0.02 $< z < $ 0.2), luminosity ($10^{9.5}\le L/L_\odot < 10^{11}$), and size (2 $\le {\rm R_{50}}/{\rm kpc} <$ 10) as candidate primary galaxies. We then compile SDSS spectroscopic lines of sight to other galaxies that are projected within 50 kpc of any candidate primary galaxy  to probe the CGM of the primary galaxy. To avoid contamination from satellites and other nearby galaxies, we require a redshift difference from the primary $>$  0.05. For each such spectrum, we fit and subtract a 10th order polynomial to a 200 \AA\ wide section at the rest wavelength, in the primary galaxy frame, for each of the emission lines of interest to remove the continuum. The emissions lines we study are 
H$\beta$, [O {\small III}] $\lambda$5007, H$\alpha$, and [N {\small II}] $\lambda$6583. 
For completeness, we also measure and detect [O {\small II}] $\lambda\lambda$3726,3729 at $> 3\sigma$ significance, but we do not discuss it further because it is not needed for the diagnostic line ratios that we discuss below.

We measure the emission flux within a prescribed velocity window relative to the primary galaxy  from each individual spectrum and combine the measurements.
For simplicity,
we set the width of that velocity window, $\pm$ 180 km s$^{-1}$, to approximate the range of kinematics expected in the halos of massive galaxies. 
This choice will include most of the absorption line systems in halos \citep{werk16}. However, we also experiment with a variable width related to an estimate of the virial velocity and describe those results, which are qualitatively similar to those obtained with the fixed window, below.
Details of how we process the data, including the rejection of outliers, are detailed in Papers I and II. 
In those papers, we adopted a broader velocity window than we do here to include as much signal as possible because were pursuing detections to large projected radii and, therefore, could not separately measure H$\alpha$ and [N{\small II}].
Finally, 
we extract M$_*$ \citep{kauffmann_agn,kauffmann_mass,Gallazzi} 
and star formation rates \citep{Brinchmann}
from the MPA-JHU catalog. \\

\section{Results}

We present measurements of the mean line fluxes of the different emission lines and the associated uncertainties, calculated using the dispersion among individual measurements, 
for lines of sight with projected radius, $r_p$, between 10 and 50 kpc in Table \ref{tab:Rdata}.
The mean $r_p$ for the lines of sight within this $r_p$ range is 35 kpc, well within the halos of most galaxies. 
Estimates of the virial radius for these galaxies, obtained as explained below, range from $\sim$ 70 to 500 kpc, although most are not as extreme and much closer to the mean, 270 kpc.

We detect statistically significant, $> 3 \sigma$, flux for all of our targeted lines in the full sample except for H$\beta$ (top line of Table \ref{tab:Rdata})\footnote{
The conversion factor to units between the values we present,  $10^{-17}$ erg cm$^{-2}$ s$^{-1}$ \AA$^{-1}$ and those used commonly in the literature to describe diffuse line emission, erg cm$^{-2}$ s$^{-1}$ arcsec$^{-2}$, is 1.7.}. 
\begin{deluxetable*}{ccccc}
\tablewidth{0pt}
\tablecaption{The emission fluxes for H$\beta$, [O III], H$\alpha$ and [N II] at different projected radius.}
\tablehead{
\colhead{$r_p$\tablenotemark{a}}  & \colhead{H$\beta$} & \colhead{[O III]} & \colhead{H$\alpha$} & \colhead{[N II]}\\ 
\colhead {[kpc]} & \multicolumn{4}{c}{[$10^{-17}$ erg cm$^{-2}$ s$^{-1}$ \AA$^{-1}$]}}
\startdata
& \multicolumn{4}{c}{Full Sample ($10^9 <$ M$_* < 10^{11.75})$} \\
35 &  $0.0017 \pm 0.0013$ & $0.0046\pm 0.0013$ & $0.0054\pm 0.0012$ & $0.0037\pm 0.0012$\\
\\
 &  \multicolumn{4}{c}{$10^9 <$ M$_* \le 10^{10.4}$ M$_\odot$} \\
 35 & $0.0015 \pm 0.0017$ & $0.0043 \pm 0.0016$ & $0.0075 \pm 0.0016$ & $<0.0014 \pm 0.0015$ \\
 17 & - &  $0.012 \pm 0.004$ & $0.022 \pm 0.004$ &  $0.0042 \pm 0.0036$ \\
 40 & - &  $0.0027 \pm 0.0018$ & $0.0045 \pm 0.0017$ &  $< 0.00054 \pm 0.0017$ \\
 \\
  &  \multicolumn{4}{c}{$10^{10.4} <$ M$_* < 10^{11.75}$ M$_\odot$} \\
  35 & $0.0019 \pm 0.0020$ & $0.0059 \pm 0.0019$ & $0.0028 \pm 0.0019$ & $0.0089 \pm 0.0018$ \\
  17 & - &  $0.024 \pm 0.005$ & $0.018 \pm 0.006$ &  $0.021 \pm 0.005$ \\
  40 & - &  $0.0026 \pm 0.0021$ & $< 0.0019 \pm 0.0019$ &  $0.0067 \pm 0.0019$
\enddata
\label{tab:Rdata}
\tablenotetext{a}{Radii refer to the mean $r_p$ of lines of sight included. A value of 35 kpc corresponds to a bin that includes all lines of sight with $r_p$ between 10 and 50 kpc. Values of 17 and 40 correspond to inner and outer $r_p$ bins within that range, respectively.}
\end{deluxetable*}

\subsection{BPT diagram}

The BPT diagnostic diagram is constructed using two line ratios, [O {\small III}]/H$\beta$ and [N {\small II}]/H$\alpha$, to define a parameter space in which one can distinguish between softer ionization sources, star formation principally, and harder sources, power law (AGN) or shocks. The use of H$\beta$ in one of the ratios and H$\alpha$ in the other is driven by the desire to avoid different extinction corrections between the numerator and denominator in those ratios. This condition is satisfied because H$\beta$ is close in wavelength to [O {\small III}] $\lambda$5007 and H$\alpha$ is close in wavelength to [N {\small II}] $\lambda$6583. 

The H$\beta$ null detection is therefore unfortunate.
The theoretical expectation for H$\beta$/H$\alpha$ has some variation that depends on physical conditions. Calculations can be done both for Case A or Case B \citep[e.g.][]{baker,hummer,osterbrock2006}, which describe whether one assumes that the gas is or is not optically thin, respectively, in the Lyman lines. Case B is typically assumed for gas in galaxies, although that gas is generally much denser than the gas in halos. Such calculations predict ratios of $\sim$ 0.35.
The H$\beta$/H$\alpha$ flux ratio we measure is $0.31\pm0.25$, which is entirely consistent
with these expectations, but is sufficiently uncertain that it provides no discriminatory power. 

Because we do not have a statistically robust determination of H$\beta$, we cannot present the standard BPT diagram. Instead, we estimate H$\beta$ using H$\alpha$ and an adopted H$\beta$/H$\alpha$ line ratio. Differential extinction between H$\alpha$ and H$\beta$ is not an issue here as extinction in galaxy halos is exceedingly low \citep{zaritsky,menard}. 
To be specific, \cite{menard} find that $A_V$ is less than 0.03 beyond projected radii of 15 kpc in their generic galaxy and is $\sim$ 0.01 at the mean $r_p$ of our sample, 35 kpc.
The use of
a correction factor between H$\alpha$ and H$\beta$ does introduce a source of uncertainty, but that can be mitigated by considering limiting cases. We choose to adopt H$\beta$/H$\alpha = 0.3$,  a round number consistent with our measurement for the entire sample and theoretical expectations. In almost all scenarios, one expects the ratio to be larger than this value and, hence, the calculated ratios may slide in one direction (downward) along the y-axis in our BPT diagrams.

\begin{figure}[htbp]
\begin{center}
\includegraphics[width = 0.48 \textwidth]{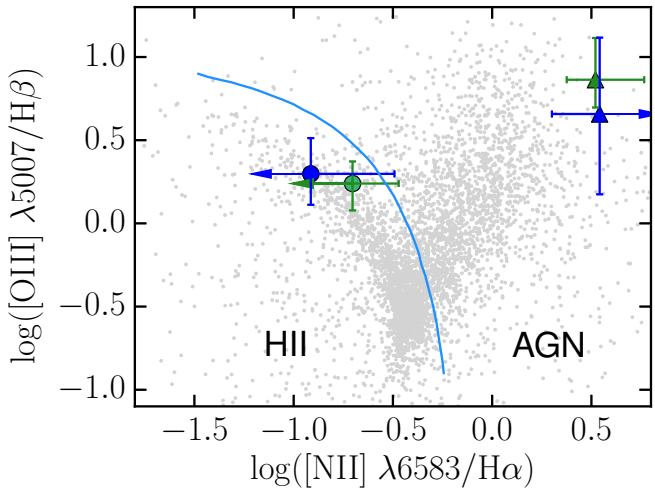}
\end{center}
\caption{The BPT diagram for circumgalactic gas within $10 < r_p/{\rm kpc} < 50$ (mean projected radius = 35 kpc; green symbols) and within a larger radius bin $22 < r_p/{\rm kpc} < 50$; blue symbols). The  circle and triangle represent the measurements for galaxies with stellar mass below and above 10$^{10.4}$ M$_\odot$, respectively.  The blue curve is the demarcation between ratios indicating ionization by star formation and AGN/shocks \citep{kauffmann_agn}, with labels indicating which region corresponds to each of the two mechanisms. The light gray points represent the line ratios for the integrated central parts of individual galaxies as measured by SDSS. The  H$\beta$ flux is estimated using the H$\alpha$ flux as described in the text.}
\label{fig:BPT_mass_radii}
\end{figure}

\begin{figure*}[htbp]
\begin{center}
\includegraphics[width = 0.75 \textwidth]{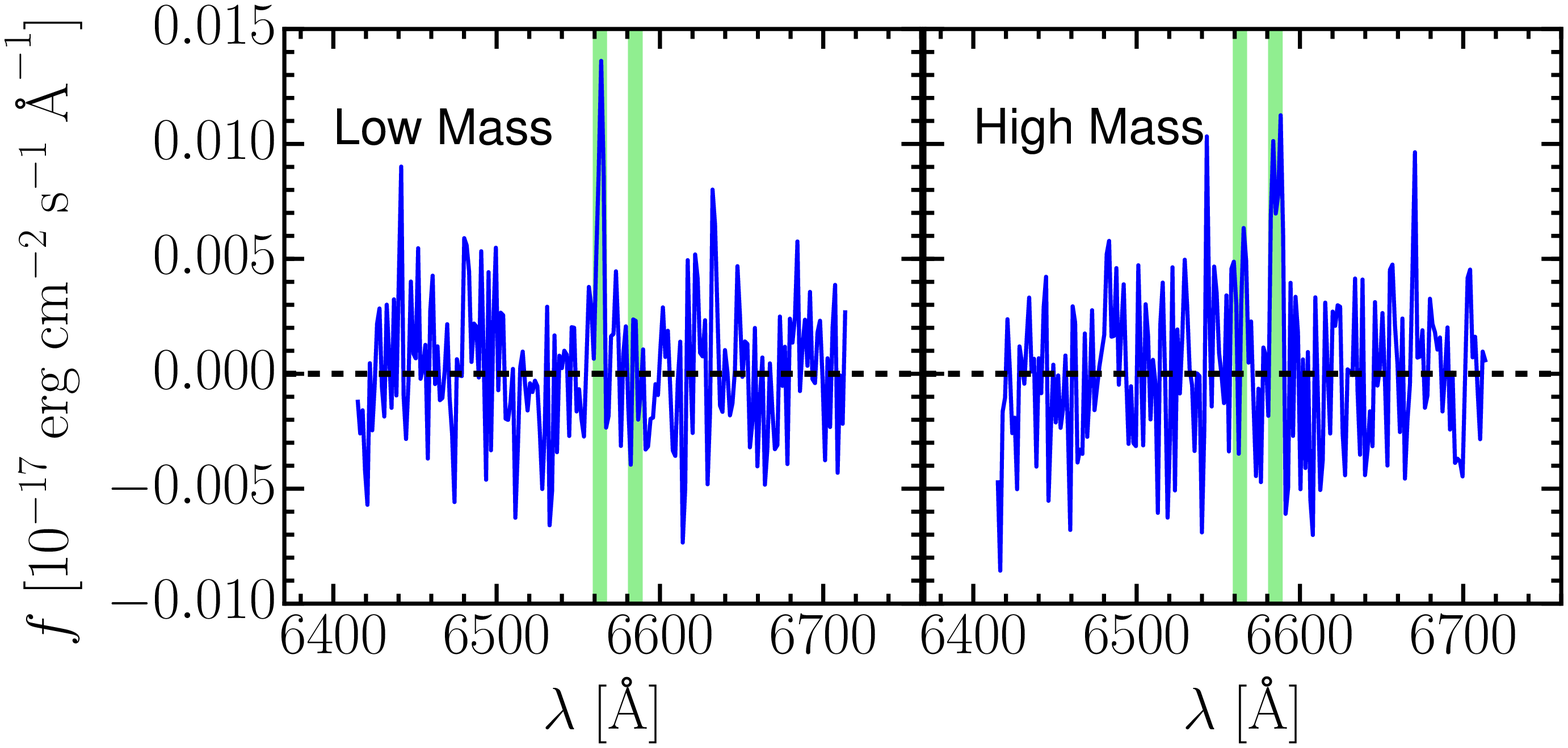}
\end{center}
\caption{The stacked composite spectra for lines of sight with 10 $<r_p <$ 50 kpc for the spectral region that includes H$\alpha$ and [N{\small II}]. The shaded regions indicate the velocity windows for those emission lines. The left panel is for the low mass subsample and the right panel is for the high mass subsample. The spectra appear noisier than our quantitative measurements because we reject outliers in our stacks only within the H$\alpha$ and N[{\small II}] windows.}
\label{fig:stack_combo}
\end{figure*}

We derive the errorbars plotted in Figure \ref{fig:BPT_mass_radii}
using a bootstrapping method. Specifically, we randomly select half of the individual measurements in each of the two mass subsamples, calculate the mean fluxes for H$\alpha$, N[{\small II}] and O[{\small III}], evaluate the ratios, treat negative ratio values as described above, and repeat the process 1000 times. We define the 16.5 and 83.5 percentiles of the resulting ratio distributions as the ends of the plotted errorbars.
The ratio values we present are the medians of the distributions.

To estimate the statistical significance of our result, we perform a bootstrap analysis by randomly splitting the sample in two, matching in size to our low and high mass subsamples. 
Among 1000 trials, we find no cases where the mean subsample line ratios match the observed values or where the lower mass sample is even farther toward the lower left in the diagram and the high mass sample farther toward the upper right, which suggests that the observational result is significant at $\ge$ 99.9\% confidence. 
For a more conservative determination, however, we measure how often we find either of the two subsamples lying on a different side of the H{\small II}/AGN demarcation line than the other subsample. Here, we find an incidence of 4\%, suggesting $>$ 95\% confidence in our result. These two estimates are likely to bracket the actual significance level.

A concern in interpreting our result is that we have mixed measurements from different radial regimes, the innermost of which may be contaminated by emission from the central galaxy. To address this concern,
we divide the data into two equal $\Delta \log r_p$ bins centered on 17 (10 $< r_p/{\rm kpc} < $ 22.4 ) and 40 kpc (22.4 $\le r_p/{\rm kpc} < $ 50). 
We present the mean flux value for each line, the uncertainty of the mean, and the average $r_p$ within each bin in Table \ref{tab:Rdata}. 
Because of the smaller sample size once we split the sample, uncertainties are larger and in two cases we measure a non-positive flux (less than 1$\sigma$ at or below 0). To enable us to present those in the log space of the BPT diagram, at least as upper limits, we added 1$\sigma$ to the measured values. The two cases where that happened are listed as upper limits in Table \ref{tab:Rdata}.
The diagnostic line ratios are consistent from the inner to outer bin in both the low and high 
stellar mass samples.  We show the larger radius bin results in Figure \ref{fig:BPT_mass_radii} for comparison to the results using the full sample.
The use of the adjusted values from the limits does not affect this conclusion because in both cases using even smaller flux values would drive the respective points farther into the area of the BPT diagram that they currently inhabit. 

To ascertain the significance of the results using only data from the outer bin, we repeat the statistical tests described previously and find that in $<$ 1\% of the cases do we achieve results as or more different than those observed and in only 13\% of the cases do the subsamples separate into the two different regions of the BPT diagram. Again, we expect these two tests to bracket the actual statistical significance. Although the statistical significance using this smaller sample is somewhat lower than that found for the entire sample, the results are entirely consistent and are now independent of any data at $r_p < 22$ kpc. We conclude that the measurements for the full sample ($10 < r_p/{\rm kpc} < 50$) can not solely be ascribed to contamination at small $r_p$, although at sufficiently small $r_p$ contamination must become a concern.

\begin{figure}[htbp]
\begin{center}
\includegraphics[width = 0.48 \textwidth]{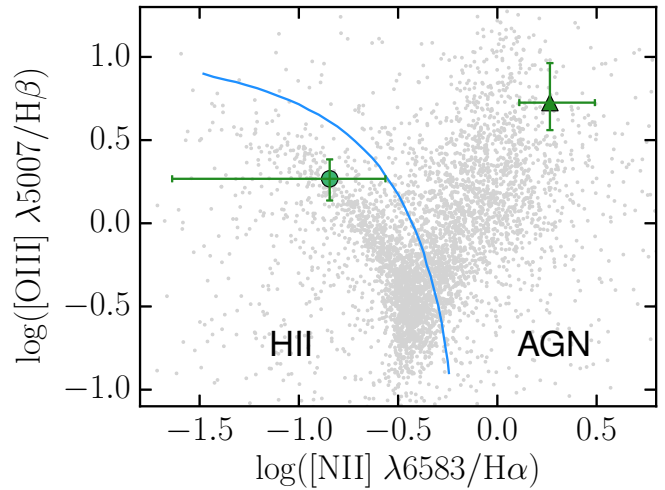}
\end{center}
\caption{
The BPT diagram for circumgalactic gas within $0.05 < r_s < 0.25$ and $r_p >$ 10 kpc. The large circle and triangle represent the measurements for galaxies with stellar mass below and above 10$^{10.4}$ M$_\odot$, respectively. The blue curve is the demarcation between ratios indicating ionization by star formation and AGN/shocks \citep{kauffmann_agn}, with labels indicating which region corresponds to each of the two mechanisms. The light gray points represent the line ratios for the integrated central parts of individual galaxies as measured by SDSS.  The H$\beta$ flux is estimated using the H$\alpha$ flux as described in the text.}
\label{fig:BPT_mass_scaleR}
\end{figure}

Another concern is that at a fixed projected radius we are probing physically different regions for galaxies with different total masses. To address this concern, we redo the analysis using a projected radius bin defined in units of scaled rather than physical radii. We set the scaled radius to be the ratio of the projected radius to the viral radius $r_s = r_p/r_\mathrm{vir}$. To estimate the halo virial radius, we use a calibration obtained from the mean relation 
between luminosity and virial radius derived using the catalog of cosmological simulations that we used in Paper II. Namely, that catalog is based on halo merger trees from the {\sl Bolshoi-Planck} simulation \citep{Klypin16,RP16}, with halos found using the \textsc{Rockstar} phase-space halo finder \citep{BehrooziRockstar} and merger trees generated with the \textsc{Consistent Trees} code \citep{BehrooziTrees}, and finally stellar masses modeled with the \textsc{UniverseMachine} code \citep{Behroozi2018}.
For a Milky Way like galaxy, the $10 < r_p/{\rm kpc} < 50$ range corresponds to $0.05 < r_s < 0.25$, and we adopt this as the range of scale  radii and require $r_p >$ 10 kpc, %{\bf
applied using the estimated virial radius of each individual galaxy.
%}. %that we use for all galaxies. {\bf 
This criterion replaces the $10 < r_p/{\rm kpc} < 50$  criterion, so the size of the sample of sightlines will differ slightly, with mean value of $r_p \sim$ 46 kpc in the range of $10 < r_p/{\rm kpc} < 136$.
In Figure \ref{fig:BPT_mass_scaleR} we present the line ratios for the stacked spectra of gas within $0.05 < r_s < 0.25$  and $r_p >$ 10 kpc, for galaxies binned by mass above and below $M_* = 10^{10.4}$ M$_\odot$. Qualitatively, the results are unchanged, although the uncertainties in these new measurements are somewhat smaller, possibly suggesting that 
there is less scatter in properties when considering scaled rather than physical radii. 

Using the same simulation data, we return to the issue of the velocity integration window. We replace the fixed width window with one of $\pm$ half the virial velocity, estimated from a scaling of the galaxy luminosity. 
The resulting average velocity windows are $\pm$80 km s$^{-1}$ and $\pm$145 km s$^{-1}$ for the low and high mass subsamples, respectively.
We find no change in the results that affects our conclusions, although the resulting measurement uncertainties are larger.

Galaxy properties are interconnected and so it is difficult to unambiguously isolate a single driver for any observed behavior. We favor mass here because of the theoretical work, but other factors that track mass, such as star formation, could be playing a role. To test for this, we set out to measure any residual correlation with star formation rate (SFR)  separately within the low and high mass samples. We find no significant differences between equally populated low ($-4 <$ log(SFR/(M$_\odot/{\rm yr})) < 1$) and high ($1 <$ log(SFR/(M$_\odot/{\rm yr})) <3)$ star formation rate subsamples populated groups within each of our two mass bins. 
In both mass bins, the high and low star formation rate samples remain consistent with the results from the full samples.

\section{Conclusions}

We present measurements of diagnostic line ratios of the line emitting warm gas in the circumgalactic medium of normal, nearby galaxies. Those line ratios indicate that lower mass galaxies, $M_* < 10^{10.4}$ M$_\odot$, have halo gas that is ionized by softer sources, similar to those found in star forming regions, while higher mass galaxies, $M_* > 10^{10.4}$ M$_\odot$, have halo gas that is ionized by harder sources, similar to found in AGN-hosting galaxies or in shocked regions. This is yet another way in which low and high stellar mass galaxies distinguish themselves from each other. 

Much of the theoretical work aiming to explain the general division of galaxy characteristics at a threshold stellar mass of $\sim 10^{10.4}$ M$_\odot$ has focused on the nature of gas accretion onto galaxies. The expectation from that work is that lower-mass galaxies have a less interrupted flow of gas to the central galaxy, leading to continual star formation, while  higher mass galaxies have inflowing gas that tends to shock and heat at large radius, interrupting the fuel flow to the central galaxy. Our observational results support this scenario, but we note the expected complexity of any full model of the baryon cycle. 

Our result is based on stacking thousands of lines of sight around thousands of nearby galaxies, and as such provides no details on the range of behavior from one galaxy to the next. However, we have empirically demonstrated that the nature of the circumgalactic gas can be explored in this manner and 
inform what would be required to do so on a case-by-case basis. Analogous
investigations will be possible for individual galaxies with the next generation of ground\peteradd{-}based large telescopes. We will be able to examine how these line ratios vary around galaxies and how they vary from galaxy to galaxy as a function of environment. Empirically tracing the nature of fuel flowing onto the central, luminous portions of galaxies will establish, refute, or necessitate revisions of what we envision is occurring at these critical scales.

\section{Acknowledgments}

DZ and HZ acknowledge financial support from NSF grant AST-1311326. JW acknowledges support from a 2018 Alfred P. Sloan Research Fellowship. The authors gratefully acknowledge  the SDSS III team for providing a valuable resource to the community.
Funding for SDSS-III has been provided by the Alfred P. Sloan Foundation, the Participating Institutions, the National Science Foundation, and the U.S. Department of Energy Office of Science. The SDSS-III web site is http://www.sdss3.org/.

SDSS-III is managed by the Astrophysical Research Consortium for the Participating Institutions of the SDSS-III Collaboration including the University of Arizona, the Brazilian Participation Group, Brookhaven National Laboratory, Carnegie Mellon University, University of Florida, the French Participation Group, the German Participation Group, Harvard University, the Instituto de Astrofisica de Canarias, the Michigan State/Notre Dame/JINA Participation Group, Johns Hopkins University, Lawrence Berkeley National Laboratory, Max Planck Institute for Astrophysics, Max Planck Institute for Extraterrestrial Physics, New Mexico State University, New York University, Ohio State University, Pennsylvania State University, University of Portsmouth, Princeton University, the Spanish Participation Group, University of Tokyo, University of Utah, Vanderbilt University, University of Virginia, University of Washington, and Yale University.

\bibliography{bibliography}

\begin{thebibliography}{42}
\expandafter\ifx\csname natexlab\endcsname\relax\def\natexlab#1{#1}\fi

\bibitem[{{Alam} {et~al}\mbox{.}(2015){Alam}, {Albareti}, {Allende Prieto},
  {Anders}, {Anderson}, {Anderton}, {Andrews}, {Armengaud}, {Aubourg},
  {Bailey}, \& et~al.}]{SDSS12}
{Alam} S. {et~al.}, 2015, \apjs, 219, 12

\bibitem[{{Angl{\'e}s-Alc{\'a}zar}
  {et~al}\mbox{.}(2017){Angl{\'e}s-Alc{\'a}zar}, {Faucher-Gigu{\`e}re},
  {Kere{\v s}}, {Hopkins}, {Quataert}, \& {Murray}}]{angles-alcazar}
{Angl{\'e}s-Alc{\'a}zar} D., {Faucher-Gigu{\`e}re} C.-A., {Kere{\v s}} D.,
  {Hopkins} P.~F., {Quataert} E., {Murray} N., 2017, \mnras, 470, 4698

\bibitem[{{Baker} \& {Menzel}(1938)}]{baker}
{Baker} J.~G., {Menzel} D.~H., 1938, \apj, 88, 52

\bibitem[{{Baldwin}, {Phillips} \& {Terlevich}(1981){Baldwin}, {Phillips}, \&
  {Terlevich}}]{bpt}
{Baldwin} J.~A., {Phillips} M.~M., {Terlevich} R., 1981, \pasp, 93, 5

\bibitem[{{Behroozi} {et~al}\mbox{.}(2018){Behroozi}, {Wechsler}, {Hearin}, \&
  {Conroy}}]{Behroozi2018}
{Behroozi} P., {Wechsler} R., {Hearin} A., {Conroy} C., 2018, ArXiv e-prints

\bibitem[{{Behroozi}, {Wechsler} \& {Wu}(2013){Behroozi}, {Wechsler}, \&
  {Wu}}]{BehrooziRockstar}
{Behroozi} P.~S., {Wechsler} R.~H., {Wu} H.-Y., 2013, \apj, 762, 109

\bibitem[{{Behroozi} {et~al}\mbox{.}(2013){Behroozi}, {Wechsler}, {Wu},
  {Busha}, {Klypin}, \& {Primack}}]{BehrooziTrees}
{Behroozi} P.~S., {Wechsler} R.~H., {Wu} H.-Y., {Busha} M.~T., {Klypin} A.~A.,
  {Primack} J.~R., 2013, \apj, 763, 18

\bibitem[{{Birnboim} \& {Dekel}(2003)}]{birnboim}
{Birnboim} Y., {Dekel} A., 2003, \mnras, 345, 349

\bibitem[{{Blanton} {et~al}\mbox{.}(2003){Blanton}, {Hogg}, {Bahcall},
  {Baldry}, {Brinkmann}, {Csabai}, {Eisenstein}, {Fukugita}, {Gunn},
  {Ivezi{\'c}}, {Lamb}, {Lupton}, {Loveday}, {Munn}, {Nichol}, {Okamura},
  {Schlegel}, {Shimasaku}, {Strauss}, {Vogeley}, \& {Weinberg}}]{blanton}
{Blanton} M.~R. {et~al.}, 2003, \apj, 594, 186

\bibitem[{{Bordoloi} {et~al}\mbox{.}(2011){Bordoloi}, {Lilly}, {Knobel},
  {Bolzonella}, {Kampczyk}, {Carollo}, {Iovino}, {Zucca}, {Contini}, {Kneib},
  {Le Fevre}, {Mainieri}, {Renzini}, {Scodeggio}, {Zamorani}, {Balestra},
  {Bardelli}, {Bongiorno}, {Caputi}, {Cucciati}, {de la Torre}, {de Ravel},
  {Garilli}, {Kova{\v c}}, {Lamareille}, {Le Borgne}, {Le Brun}, {Maier},
  {Mignoli}, {Pello}, {Peng}, {Perez Montero}, {Presotto}, {Scarlata},
  {Silverman}, {Tanaka}, {Tasca}, {Tresse}, {Vergani}, {Barnes}, {Cappi},
  {Cimatti}, {Coppa}, {Diener}, {Franzetti}, {Koekemoer}, {L{\'o}pez-Sanjuan},
  {McCracken}, {Moresco}, {Nair}, {Oesch}, {Pozzetti}, \&
  {Welikala}}]{bordoloi2011}
{Bordoloi} R. {et~al.}, 2011, \apj, 743, 10

\bibitem[{{Brinchmann} {et~al}\mbox{.}(2004){Brinchmann}, {Charlot}, {White},
  {Tremonti}, {Kauffmann}, {Heckman}, \& {Brinkmann}}]{Brinchmann}
{Brinchmann} J., {Charlot} S., {White} S.~D.~M., {Tremonti} C., {Kauffmann} G.,
  {Heckman} T., {Brinkmann} J., 2004, MNRAS, 351, 1151

\bibitem[{{Croft} {et~al}\mbox{.}(2018){Croft}, {Miralda-Escud{\'e}}, {Zheng},
  {Blomqvist}, \& {Pieri}}]{croft2018}
{Croft} R.~A.~C., {Miralda-Escud{\'e}} J., {Zheng} Z., {Blomqvist} M., {Pieri}
  M., 2018, \mnras

\bibitem[{{Croft} {et~al}\mbox{.}(2016){Croft}, {Miralda-Escud{\'e}}, {Zheng},
  {Bolton}, {Dawson}, {Peterson}, {York}, {Eisenstein}, {Brinkmann},
  {Brownstein}, {Cen}, {Delubac}, {Font-Ribera}, {Hamilton}, {Lee}, {Myers},
  {Palanque-Delabrouille}, {P{\^a}ris}, {Petitjean}, {Pieri}, {Ross}, {Rossi},
  {Schlegel}, {Schneider}, {Slosar}, {Vazquez}, {Viel}, {Weinberg}, \&
  {Y{\`e}che}}]{croft2016}
{Croft} R.~A.~C. {et~al.}, 2016, \mnras, 457, 3541

\bibitem[{{Dekel} \& {Birnboim}(2006)}]{dekel}
{Dekel} A., {Birnboim} Y., 2006, \mnras, 368, 2

\bibitem[{{Ford} {et~al}\mbox{.}(2014){Ford}, {Dav{\'e}}, {Oppenheimer},
  {Katz}, {Kollmeier}, {Thompson}, \& {Weinberg}}]{ford}
{Ford} A.~B., {Dav{\'e}} R., {Oppenheimer} B.~D., {Katz} N., {Kollmeier} J.~A.,
  {Thompson} R., {Weinberg} D.~H., 2014, \mnras, 444, 1260

\bibitem[{{Gallazzi} {et~al}\mbox{.}(2005){Gallazzi}, {Charlot}, {Brinchmann},
  {White}, \& {Tremonti}}]{Gallazzi}
{Gallazzi} A., {Charlot} S., {Brinchmann} J., {White} S.~D.~M., {Tremonti}
  C.~A., 2005, MNRAS, 362, 41

\bibitem[{{Hummer} \& {Storey}(1987)}]{hummer}
{Hummer} D.~G., {Storey} P.~J., 1987, \mnras, 224, 801

\bibitem[{{Joshi} {et~al}\mbox{.}(2018){Joshi}, {Srianand}, {Petitjean}, \&
  {Noterdaeme}}]{joshi2018}
{Joshi} R., {Srianand} R., {Petitjean} P., {Noterdaeme} P., 2018, \mnras, 476,
  210

\bibitem[{{Kauffmann} {et~al}\mbox{.}(2003{\natexlab{a}}){Kauffmann},
  {Heckman}, {Tremonti}, {Brinchmann}, {Charlot}, {White}, {Ridgway},
  {Brinkmann}, {Fukugita}, {Hall}, {Ivezi{\'c}}, {Richards}, \&
  {Schneider}}]{kauffmann_agn}
{Kauffmann} G. {et~al.}, 2003{\natexlab{a}}, \mnras, 346, 1055

\bibitem[{{Kauffmann} {et~al}\mbox{.}(2003{\natexlab{b}}){Kauffmann},
  {Heckman}, {White}, {Charlot}, {Tremonti}, {Peng}, {Seibert}, {Brinkmann},
  {Nichol}, {SubbaRao}, \& {York}}]{kauffmann_mass}
{Kauffmann} G. {et~al.}, 2003{\natexlab{b}}, \mnras, 341, 54

\bibitem[{{Kere{\v s}} {et~al}\mbox{.}(2005){Kere{\v s}}, {Katz}, {Weinberg},
  \& {Dav{\'e}}}]{keres}
{Kere{\v s}} D., {Katz} N., {Weinberg} D.~H., {Dav{\'e}} R., 2005, \mnras, 363,
  2

\bibitem[{{Kewley} {et~al}\mbox{.}(2001){Kewley}, {Dopita}, {Sutherland},
  {Heisler}, \& {Trevena}}]{kewley}
{Kewley} L.~J., {Dopita} M.~A., {Sutherland} R.~S., {Heisler} C.~A., {Trevena}
  J., 2001, \apj, 556, 121

\bibitem[{{Klypin} {et~al}\mbox{.}(2016){Klypin}, {Yepes}, {Gottl{\"o}ber},
  {Prada}, \& {He{\ss}}}]{Klypin16}
{Klypin} A., {Yepes} G., {Gottl{\"o}ber} S., {Prada} F., {He{\ss}} S., 2016,
  \mnras, 457, 4340

\bibitem[{{Lan} \& {Mo}(2018)}]{lan2018}
{Lan} T.-W., {Mo} H., 2018, ArXiv e-prints

\bibitem[{{M{\'e}nard} {et~al}\mbox{.}(2010){M{\'e}nard}, {Scranton},
  {Fukugita}, \& {Richards}}]{menard}
{M{\'e}nard} B., {Scranton} R., {Fukugita} M., {Richards} G., 2010, \mnras,
  405, 1025

\bibitem[{{M{\'e}nard} {et~al}\mbox{.}(2011){M{\'e}nard}, {Wild}, {Nestor},
  {Quider}, {Zibetti}, {Rao}, \& {Turnshek}}]{menard2011}
{M{\'e}nard} B., {Wild} V., {Nestor} D., {Quider} A., {Zibetti} S., {Rao} S.,
  {Turnshek} D., 2011, \mnras, 417, 801

\bibitem[{{Oppenheimer} {et~al}\mbox{.}(2018){Oppenheimer}, {Segers}, {Schaye},
  {Richings}, \& {Crain}}]{oppenheimer}
{Oppenheimer} B.~D., {Segers} M., {Schaye} J., {Richings} A.~J., {Crain} R.~A.,
  2018, \mnras, 474, 4740

\bibitem[{{Osterbrock} \& {Ferland}(2006)}]{osterbrock2006}
{Osterbrock} D.~E., {Ferland} G.~J., 2006, {Astrophysics of gaseous nebulae and
  active galactic nuclei}

\bibitem[{{Prochaska} {et~al}\mbox{.}(2017){Prochaska}, {Werk}, {Worseck},
  {Tripp}, {Tumlinson}, {Burchett}, {Fox}, {Fumagalli}, {Lehner}, {Peeples}, \&
  {Tejos}}]{prochaska2017}
{Prochaska} J.~X. {et~al.}, 2017, \apj, 837, 169

\bibitem[{{Roberts} \& {Haynes}(1994)}]{roberts}
{Roberts} M.~S., {Haynes} M.~P., 1994, \araa, 32, 115

\bibitem[{{Rodr{\'{\i}}guez-Puebla}
  {et~al}\mbox{.}(2016){Rodr{\'{\i}}guez-Puebla}, {Behroozi}, {Primack},
  {Klypin}, {Lee}, \& {Hellinger}}]{RP16}
{Rodr{\'{\i}}guez-Puebla} A., {Behroozi} P., {Primack} J., {Klypin} A., {Lee}
  C., {Hellinger} D., 2016, \mnras, 462, 893

\bibitem[{{Steidel} {et~al}\mbox{.}(2010){Steidel}, {Erb}, {Shapley},
  {Pettini}, {Reddy}, {Bogosavljevi{\'c}}, {Rudie}, \& {Rakic}}]{steidel2010}
{Steidel} C.~C., {Erb} D.~K., {Shapley} A.~E., {Pettini} M., {Reddy} N.,
  {Bogosavljevi{\'c}} M., {Rudie} G.~C., {Rakic} O., 2010, \apj, 717, 289

\bibitem[{{Tumlinson}, {Peeples} \& {Werk}(2017){Tumlinson}, {Peeples}, \&
  {Werk}}]{tumlinson}
{Tumlinson} J., {Peeples} M.~S., {Werk} J.~K., 2017, \araa, 55, 389

\bibitem[{{Veilleux} \& {Osterbrock}(1987)}]{vo}
{Veilleux} S., {Osterbrock} D.~E., 1987, \apjs, 63, 295

\bibitem[{{Werk} {et~al}\mbox{.}(2016){Werk}, {Prochaska}, {Cantalupo}, {Fox},
  {Oppenheimer}, {Tumlinson}, {Tripp}, {Lehner}, \& {McQuinn}}]{werk16}
{Werk} J.~K. {et~al.}, 2016, \apj, 833, 54

\bibitem[{{Werk} {et~al}\mbox{.}(2014){Werk}, {Prochaska}, {Tumlinson},
  {Peeples}, {Tripp}, {Fox}, {Lehner}, {Thom}, {O'Meara}, {Ford}, {Bordoloi},
  {Katz}, {Tejos}, {Oppenheimer}, {Dav{\'e}}, \& {Weinberg}}]{Werk2014}
{Werk} J.~K. {et~al.}, 2014, \apj, 792, 8

\bibitem[{{White} \& {Rees}(1978)}]{white}
{White} S.~D.~M., {Rees} M.~J., 1978, \mnras, 183, 341

\bibitem[{{Zaritsky}(1994)}]{zaritsky}
{Zaritsky} D., 1994, \aj, 108, 1619

\bibitem[{{Zhang}, {Zaritsky} \& {Behroozi}(2018){Zhang}, {Zaritsky}, \&
  {Behroozi}}]{zhang2018}
{Zhang} H., {Zaritsky} D., {Behroozi} P., 2018, \apj, 861, 34

\bibitem[{{Zhang} {et~al}\mbox{.}(2016){Zhang}, {Zaritsky}, {Zhu},
  {M{\'e}nard}, \& {Hogg}}]{zhang2016}
{Zhang} H., {Zaritsky} D., {Zhu} G., {M{\'e}nard} B., {Hogg} D.~W., 2016, \apj,
  833, 276

\bibitem[{{Zhu} \& {M{\'e}nard}(2013{\natexlab{a}})}]{zhu2013a}
{Zhu} G., {M{\'e}nard} B., 2013{\natexlab{a}}, \apj, 773, 16

\bibitem[{{Zhu} \& {M{\'e}nard}(2013{\natexlab{b}})}]{zhu2013b}
{Zhu} G., {M{\'e}nard} B., 2013{\natexlab{b}}, \apj, 770, 130

\end{thebibliography}

\end{document}